\documentclass[conference]{IEEEtran}
\usepackage{blindtext, graphicx}

\usepackage{algorithm}
\usepackage{algpseudocode}

\usepackage{graphicx} 
\graphicspath{ {} }

\begin{document}
%
\title{Assisting humans to achieve optimal sleep by changing ambient temperature}

\author{\IEEEauthorblockN{Vivek Gupta*}
\IEEEauthorblockA{IIT Kanpur\\
keviv9@gmail.com}
\and
\IEEEauthorblockN{Siddhant Mittal*}
\IEEEauthorblockA{IIT Bombay\\
siddhantiitbmittal3@gmail.com}
\and
\IEEEauthorblockN{Sandip Bhaumik}
\IEEEauthorblockA{Samsung R\&D,Bangalore\\
sandip.bh@samsung.com}
\and
\IEEEauthorblockN{Raj Roy}
\IEEEauthorblockA{Samsung R\&D,Bangalore\\
raj.roy@samsung.com}}


\maketitle

\begin{abstract}
\footnote{*Equal and Major Contribution}
Environment plays a vital role in the sleep mechanism of a human. It has been shown from many studies that sleeping and waking environment, waking time and hours of sleep is of very significant importance \cite{J Sleep Res} which can result in sleeping disorders and variety of diseases. This paper finds the sleep cycle of an individual and according changes the ambient temperature to maximize his/her sleep efficiency. We suggest a method which will assist in increasing sleep efficiency. Using Fast-Fourier-Transformation (FFT) of heart rate signals to extract heart rate variability data such that low frequency / high frequency (LF/HF) power ratio we are detecting sleep stages using an automated algorithm and then applying feedback mechanism to alter the ambient temperature depending upon the sleep stage.
\end{abstract}

\begin{IEEEkeywords}
Heart Rate Variability (HRV), Sleep cycle, Sleep stages, ambient temperature.
\end{IEEEkeywords}

%
\IEEEpeerreviewmaketitle

\section{Introduction}
Noctural sleep is divided into cycles. Each cycle is repeated on an average 4-5 times during night in an healthy individual, and each cycle contains a cascade of sleep stages normally in the same order of occurrence(but in different proportion in the night). Sleep stages are determine according to a set of rules defined by Rechtschaffen and Kales(R\&K) \cite{RandK} It is accustomed to defined 3 stages of existence: wake REM sleep and Non-REM sleep. The later is further subdivided into Light sleep i.e. LS (stage I and stage 2) and slow wave sleep i.e. SWS (stage 3 and stage 4). \cite{Heart Rate Spectrum Analysis}

The sleep detection of a sleep is done in laboratory using an expensive and uncomfortable sleep study which involves the recording of various bio-signals including EEG, EOG, EMG, ECG, pulse oximetry and various breathing related signals. In this research we have used an alternative method for detection of sleep using ECG signals alone because it alone contains relevant information regarding sleep (e.g sleep onset and arousals from sleep during the night), sleep disturbance and sleep structures (sleep stages). The information regarding sleep structure can be uncovered by spectrum analysis of the inter-beat (RR) interval signal. Studies have shown the High Frequency (HF) percentage of total power is increased during SWS with the decrease in Low Frequency (LF) power band. Also, studies have shown the ratio of LF/HF is more than unity during night but during SWS it is less than unity. 

Using Spectral Analysis techniques like Heart Rate Variability (HRV) we have consistently found increased parasympathetic activity during NREM sleep as measured by high frequency heart rate variability and increased sympathetic activity during REM sleep, as typically measured by the ratio of lower to higher frequency heart rate variability. These findings agree well with earlier studies of heart rate and direct measure of nerve activity. There is an increase in parasympathetic activities during stage 2 (light sleep) and decrease of both these activities in SWS and increase of sympathetic activities in REM stage.

Findings indicate that maintaining a comfortable thermal sleep environment is important for sleep as well as daytime activities and health status. Humans have a sleep-wake rhythm that is repeated in a 24-hour cycle. The core body temperature (Tcore), which also cycles along with the sleep-wake rhythm, decreases during the nocturnal sleep phase and increases during the wake phase repeatedly in a 24-hours circadian rhythm. Sleep is most likely to occur when Tcore decreases. The relationship between the sleep-wake rhythm and the circadian rhythm of Tcore is important for maintaining sleep. \cite{Effects of thermal}

On the sleep onset our Tcore decreases due to low metabolism rate of the body. During this time the ambient temperature has to be low to facilitate rapid heat loss. After sleep onset, Tcore gradually decreases further, while distal and proximal Tsk (skin temperature) remain high, even a slight increase in proximal Tsk increases the amount of SWS and decrease the early-morning awakening in the elderly.

Thermoregulatory response during sleep differs depending on sleep stages. Sensitivity to hot and cold stimulation is reduced in REM compared to non-REM and wakefulness. Sweat rate increases during SWS compared to other sleep stages whereas sweat rate and sweating decreases in REM. Thermoregulation and REM are mutually exclusive and partly explain the decrease in REM observed during heat and cold exposure telling us REM is more sensitive to Ta (ambient temperature).

Taking these considerations into account we developed an automated algorithm with uses spectral analysis of heart rate to detect sleep and apply a feedback mechanism to alter ambient temperature. Although this paper gives many insights into how to we sleep and how we can improve sleep efficiency but the two main contribution of this paper is as follows

\begin{itemize}
\item Firstly, this paper is first of it's kind which relates how the surrounding (temperature in our case) affects our sleep.

\item Secondly, it develops a novel personalised approach to improve sleep efficiency at maximum comfort level.

\end{itemize}

\section{Theoretical Background}
Sleep is very important for proper functioning of the human body. Sleeping is the time when our body goes into repairing, maintained and growth activities of our body \cite{Sleep patterns}. It has a direct effect on our circadian cycle. Furthermore, disturbed nocturnal sleep affects not only daytime activities, but also leads to various adverse health effects, such as obesity \cite{association between short sleep}, quality of life and even mortality \cite{Mechanism and functions,Healthy older}. Sleeping is divided into many sleep cycle, each sleep cycle having 4 (or 5 stages depending upon nomenclature) first 3 states are stage 1, stage 2, and Slow Wave Sleep (SWS) also known as deep sleep. The 4th stage is known as Rapid Eye moment (REM) stage. In a cycle, rectral (core body) temperature, skin temperature, heart rate, brain activities, blood pressure also changes significantly in each stage. Good sleep has about 4-6 cycles depending upon person to person. In infants sleeping time adequate for the body is 12-14 hours and in adults it is 7-8 hours. In infants REM proportion is higher than adults.

REM stage is important for mental and memory recovery and learning whereas SWS is important for repair, immune system and biological process, it is also known as deep sleep and brain is least active this time. Each Sleep cycle is approx. about 90 $\pm$ 20 minutes.

There are two type of episode in sleeping

\begin{itemize}
\item Single Sleep episode:- Single Sleep episode is completing your sleeping process without any break in cycles. This is an ideal situation and is required for proper intervals of SWS and REM sleep.

\item Across Sleep episode:- Across Sleep episode is breaking of sleeping cycles due to arousals. Studies has shown that renewal of cycles occur but important Stages like SWS and REM are suppressed in coming up phases, this is the case with people having apnea \cite{The REM-NREM sleep cycle}
\end{itemize}

We will be using heart rate variability to recognize the sleep cycle of an individual. Heart rate is also related to sleep stage, awakenings and body movement. Best way of detecting someones sleep stage is by monitoring brain activities. The results replicated previous studies in showing increases in high frequency components and decreases in low frequency components of heart rate variability across NREM sleep stages and opposite changes in REM sleep and wake. These results are consistent with sympathetic nervous system activation during REM sleep and wake periods. The shift in heart rate variability seen during REM sleep began in NREM sleep several minutes prior to standardly scored REM and often continued beyond the end of REM sleep \cite{Age related modification of NREM sleep}

Two important activities of autonomous nervous system 
\begin{enumerate}
  \item Sympathetic Activities :- Quick response, mobilize the body's nervous system fight-or-flight response
  \item Parasympathetic Activities :- Biological processes
such as digestion, salivation, urination etc
\end{enumerate}

Using Spectral Analysis techniques (HRV) we have consistently found increased parasympathetic activity during NREM sleep as measured by high frequency heart rate variability and increased sympathetic activity during REM sleep, as typically measured by the ratio of lower to higher frequency heart rate variability. These findings agree well with earlier studies of heart rate and direct measure of nerve activity. There is an increase in parasympathetic activities during stage 2 and decrease of both these activities in SWS and increase of sympathetic activities in REM stage. After getting a sleep cycle we locally optimize the environment conditions to enhance efficiency.

\subsection{Heat Exposure}
\begin{itemize}
\item Finding has suggested that wakefulness is the only stage which can cope with an increased thermal load \cite{The Human Circadian Timing System and Sleep-Wake Regulation} and that wakefulness replaces SWS and REM to maintain homoeothermy.

\item Heat exposures during onset on sleep can suppress the decrease in Tcore, increasing Tsk.

\item Humid	heat exposure further increases 
wakefulness, decreases REM and SWS and excessively suppresses the decrease in Tcore. whereas Tsk and whole body sweat loss are not affected.
\end{itemize}

\subsection{Cold Exposure}
\begin{itemize}
  \item The difference between cold exposure and heat exposure is that cold exposure mainly affects the later segments of sleep
  \item Cold exposure mostly affects REM due to lack of thermoregulatory
  \item Results have shown that parasympathetic activities increase with low ambient temperature(Ta), hence we need a low Ta in Stage 1 and 2
  \item Since 70 \% of sleep is spent in Stage 1, 2 and SWS thus cold ambient temperature are preferred. It is extremely important to note that Ta in winter should be maintained at a level higher than 10$^{\circ}$ C, However, the most difficult aspect of cold exposure is that sleep is not disturbed. The impact on the cardiovascular response may he occurring without subjective sensation, suggesting that cold exposure may more impact than heat exposure.
\end{itemize}

\section{Methods}
Description of Algorithm

\subsection{Step 1 : Frequency Domain analysis}
By Heart rate spectral analysis, we analyse usually duet band waves (i) VLF (very low frequency) (0-0.04 Hz) (ii) LF (low frequency) (0.04 — 0.15 Hz) and (iii) HF thigh frequency (0.15 — 0.4 Hz). We convert the heart beat per minutes into RR interval which is the time between two beats. Using Fast-Fourier-transformation (FFT) and frequency domain analysis (FDA) the three band frequency VLF. LF and HF is detected on an interval of 5 minutes.

\subsection{Step 2 : Sleep stage detection}
Sleep stages are detected using the spectrum analysis in step 1. LF band value is increased in REM and wakefulness and HF value is increased during SWS. Both band power spectra are high in wakefulness. Studies have shown that LF/HF ratio is greater than unity at night except for SWS stages. Since sleep stages occur normally in the same order of occurrence then stage 3 will follow stage 4 in SWS stages when we find that LF/HF is less than unit \cite{Automatic detection}. For differentiating of sleep stages from calculated 
signals LF/HF frequency power ratio and relative peak frequency-power within HF band, threshold value for both signals hare to be equated and optimized. Which can be done by generic optimizing algorithm which can he applied on learning set, which is not the current part of the algorithm of this research paper and can be considered as future work. For the variability of the peak frequency power we applied a fixed threshold value of 0.65 Hz. as no erroneous classification of NREM sleep could be observed below this variability \cite{Heart Rate Spectrum Analysis}.

For sleep detection first whole recording was identified as light sleep (NREM 1-2) as more than 50 \% sleep is spent in light sleep. LF/HF ratio is used to identify deep sleep. Last step is distinguishing REM, for this HF power greater than 0.65 Hz was consider as REM sleep.

\subsection{Step 3: Feedback loop Mechanism}
 We used regular interval window to detect the stage and give feedback. If the system detect that sleep is in transition state then feedback loop is important because it will decide the amount to be changed in temperature. In every window we decrease or increase the temperature accordingly by delta amount.  Advantage of feedback loop is including the subject’s interference. The temperature modulation will occur around the subject's provided external temperature so as his/her comfortability is given priority.Algorithm creates a Deterministic Finite Automata (Fig 1) having 3 states i.e. State Plus, State Minus and State Neutral. 

\includegraphics[width=7.5cm, height=6.5cm]{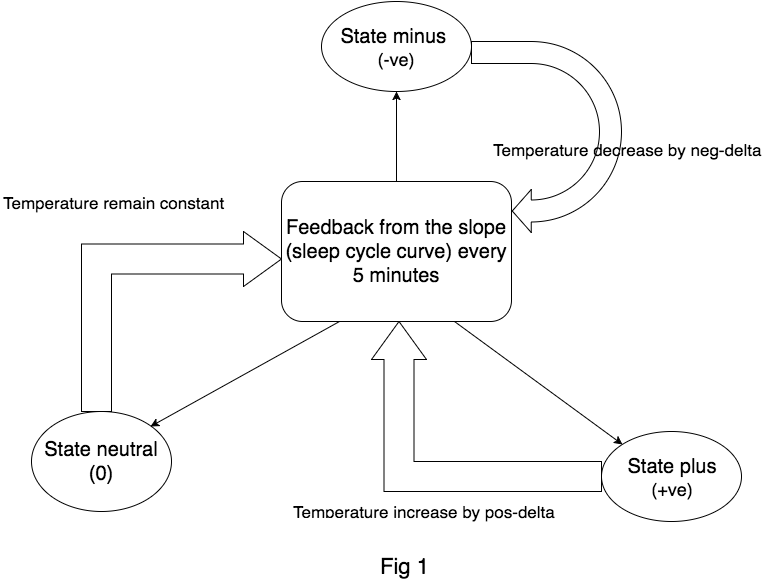}

If the slope last window (Sleep Cycle) is negative, we stay in State Minus and decrease temperature by neg-delta. If the slope of last window (Sleep Cycle) is positive we stay in State Plus and increase temperature by pos-delta, otherwise we stay in State Neutral and keep temperature constant. The state is checked in same window time (5 min) interval continuously. A state change logic is implemented which measure slope of last window and switch state if required. Hence ambient temperature is feed backed from sleep cycle states in periodic cycles of 5 minutes

\begin{algorithm}
\caption{Algorithm for Feedback mechanism}
\label{Algorithm for Feedback mechanism}
1.	Ambient temperature = as specified by user \newline
2.	Deterministic state = 0 \newline
3.	\textbf {state\_minis()} \newline
\hspace*{10 mm} a.	deterministic state = -1 \newline
\hspace*{10 mm} b.	return -negdeltha \newline
4.  \textbf {state\_plus()} \newline
\hspace*{10 mm} a.	deterministic state = 1 \newline
\hspace*{10 mm} b.	return posdeltha \newline
5.	\textbf {state\_neutral()} \newline
\hspace*{10 mm} a.	deterministic state = 0 \newline
\hspace*{10 mm} b.	return 0 \newline
6.	\textbf {Feedbackloop} \newline
Input: current state, previous state \newline
Output: deterministic state \newline
\hspace*{10 mm} a.	Slope = (current state - previous state)  \newline
\hspace*{10 mm} b.	if slope positive  \newline
\hspace*{10 mm} \hspace*{10 mm} Deterministic state is increased  \newline
\hspace*{10 mm} \hspace*{10 mm} state\_plus()  \newline
\hspace*{10 mm} \hspace*{10 mm} Temperature profile ← positive delta  \newline
\hspace*{10 mm} c.	if slope negative  \newline
\hspace*{10 mm} \hspace*{10 mm} Deterministic state is decreased  \newline
\hspace*{10 mm} \hspace*{10 mm} State\_minus()  \newline
\hspace*{10 mm} \hspace*{10 mm} Temperature profile ← negative delta  \newline
\hspace*{10 mm} d.	if slope zero  \newline
\hspace*{10 mm} \hspace*{10 mm} Deterministic state is kept unchanged  \newline
\hspace*{10 mm} \hspace*{10 mm} state\_neutal()  \newline
\hspace*{10 mm} \hspace*{10 mm} Temperature profile ←  0  \newline
7.	\textbf {Temperature profile}  \newline
Input:  delta  \newline
\hspace*{10 mm} a.	Ambient temperature += detla  \newline
8.	\textbf{Main loop}  \newline
\hspace*{10 mm} a.	while (no more sleep stage to detect)  \newline
\hspace*{10 mm} \hspace*{10 mm} Previous state = current state  \newline
\hspace*{10 mm} \hspace*{10 mm} Current state = Getnextsleepstate()  \newline
\hspace*{10 mm} \hspace*{10 mm}Feedbackloop (current state, previous state, deterministic state)  \newline
\end{algorithm}

\subsection{Step 4: Temperature modulation}
If the current slope of sleep cycle is moving toward Deep Sleep i.e. -ve slope then we decrease the temperature by neg¬-delta amount (set accordingly). If the current slope of sleep cycle is moving toward REM stage i.e. +ve slope then we decrease the Temperature by pos-delta amount (set accordingly), Initially neg-delta is of higher value (set to 0.6) and pos-delta value (set to 0.4) because it require lower temperature during onset of temperature. Also during initial phase NREM phase is high and hence neg-delta should he high (decrease in temperature). Since REM stages are more prominent in later phase value of pos-delta increase and after 3-4 hours (set accordingly) exceed neg-delta. We decrease the temperature difference between (neg-delta - pos-delta) which represent the increase and decrease parameter of temperature. With each cycle this difference is reduced with time.

Feedback loop will decide the state of external parameter. For example, if the slope of the sleep cycle is -ve then we will decrease the Ambient Temperature (Ta). The external control will be with both the method and subject. The external control only controls the current decrease (- ve) or increase (+ve) or no increase or decrease in the temperature.

\section{Result and Observation}
For simulation and result, automated script was created which took RR interval as its input. RR is the beat to beat time taken in seconds. Spectrum analysis is done by taking a window of 5 minutes. In sleep detection we have increased the window to 15 minutes by taking the average of these three 5 minute windows.

Fig 2 shows the spectrum analysis of heart rate corresponding to RR value recorded during a particular section of sleep time. In this figure we can see the black circle showing the detection of deep sleep stages. As the LF/HF ratio decreases below unity we enter into the deep sleep stages. Similarly, decreased HF value indicates entering into the REM stage.

\includegraphics[width=8cm, height=10cm]{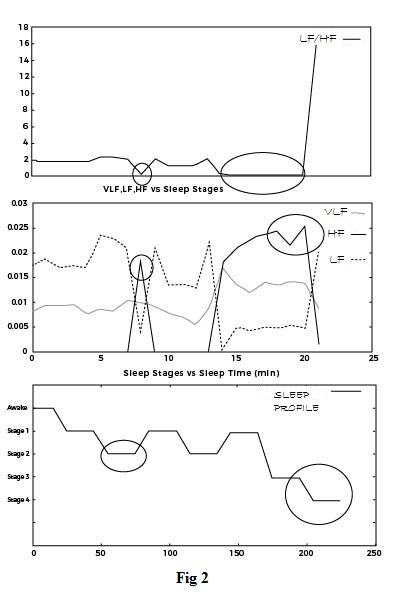}

Above result matches with the theoretical background behind frequency analysis and sleep stage detection.

Fig 3 is the sleep cycle detected after step 2 of the algorithm, it shows the sleep stages vs sleep time and Fig 4 shows corresponding variation of ambient temperature. This temperature modulation occurs after applying the step 4 and step 5 of the algorithm.

\includegraphics[width=8cm, height=6cm]{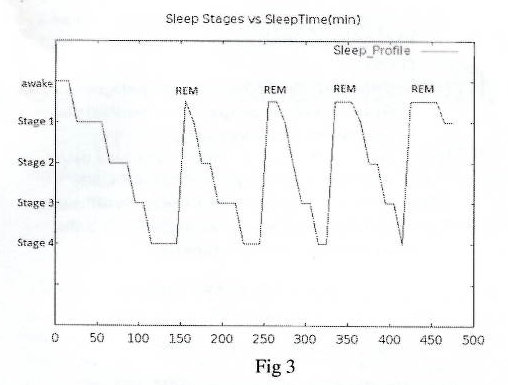}

\includegraphics[width=8cm, height=6cm]{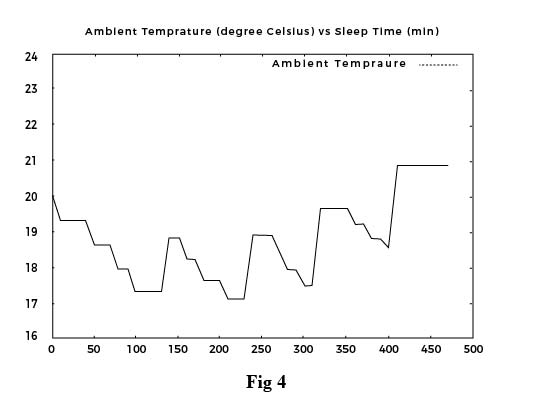}

Fig 5 shows the percentage wise time spent in each stage. This will be personalized depending upon the subject under consideration because time spend in each stage depends upon the age, sex and physical fitness of the subject.

\includegraphics[width=8cm, height=6cm]{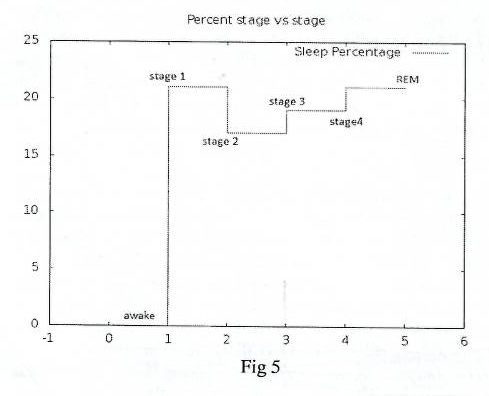}

\section{Conclusion}
We successfully developed an algorithm to provide healthy personalized sleep. Our methods will assist the user to achieve sleep by detecting the sleep cycles and using it to improve sleep stages, avoiding sleep fragmentation and improving biological activities depending upon the sleep stage the user is in. Our method will adjust the ambient conditions depending upon sleep cycle. Another benefit of this method is the personification of the systems. Our methods is taking user interference into account so user’s comfortability come first and thus making each system tuned to its user rather than providing generalized pattern for everybody.

Despite the fact that we are using a rough approximation for sleep detection which largely depends upon our selected window and subject, this is best approach for sleep detection using ECG according to current state of art (not using clinical routine).
With the results obtained from our algorithm we can see variation in ambient temperature while sleeping avoid sleepiness in daytime and also maintain proper wake sleep rhythm. In combination with the other algorithm using ECG signals like e.g. for the detection of sleep related breathing disorders and sleep detection depending upon body movements etc., \textbf{we believe our approach represents the first step towards this idea of changing ambient environment to provide healthy sleep}.

Compared to current state of Art currently to get a sleep cycle detection there are either two approaches one is clinical and the other is industrial.

Clinical approach involves lot of heavy and costly machines, which are uncomfortable to wear while sleeping. This approach uses ECG, brain activities, body movement, and blood pressure to detect sleep cycle. Our paper gives a novel and practical approach towards sleep detection and improve sleep efficiency.

The industrial approach which includes lot of application out in the market which either uses your body movement, breathing sounds etc to detect sleep cycle are not very accurate as it lacks theoretical background. Our approach has a strong biological and technical background. Even though our approach and algorithm as of now is a plug and play approach and more practical it has a lot of future development prospective of learning human sleep behavior and applying deep learning to improve sleep efficiency.

Lastly authors of this paper believe in serving mankind and would also like to use this novel approach more clinically in biomedicine to tackle sleep diseases.

\section*{Acknowledgment}
The authors will like to thank Physionet HRV toolkit (GNU general public license) for helping us understand the frequency domain analysis of Heart rate signals and using in our paper.
Work was done by the first and the second authors equally when they were research intern at Samsung R\&D Institute Bangalore.

%

\begin{IEEEbiography}[{\includegraphics[width=1in,height=1.25in,clip,keepaspectratio]{picture}}]{John Doe}
\blindtext
\end{IEEEbiography}




\end{document}